\documentclass{PoS}
\usepackage{amsmath}
\title{Anomalous processes and leading logarithms}

\ShortTitle{Anomalous processes and leading logs}

\author{\speaker{Karol Kampf}\thanks{I would like to thank the conference
organizers for an enjoyable meeting and my
collaborators I had the pleasure to work with on different parts of this
report: Johan Bijnens, Marc Knecht, Stefan Lanz, Bachir Moussallam and
Ji\v{r}\'{\i} Novotn\'{y}. This work is supported in part by MSM0021620859 of
the Ministry of Education of the Czech Republic and by the project UNCE no.
204020/2012 of Charles University.}\\
        Institute of Particle and Nuclear Physics, 
	Faculty of Mathematics and Physics,\\
	Charles University, CZ-18000 Prague, Czech Republic.\\
        E-mail: \email{karol.kampf@mff.cuni.cz}}

\abstract{Present status of odd-intrinsic sector of low energy QCD is
summarized. The two-photon decay of neutral pion is shortly discussed and its
connection with the pion decay constant is analysed. A theoretical tool, the 
leading-log calculation is also presented, its connection with anomalous
sector shown and some new results given.}

\FullConference{The 7th International Workshop on Chiral Dynamics,\\
		August 6 -10, 2012\\
		Jefferson Lab, Newport News, Virginia, USA}

\begin{document}

\section{Introduction}
Chiral perturbation theory (ChPT) represents a successful effective field theory
of quantum chromodynamics (QCD). For slightly more detailed discussion on
its present status and for references see dedicated talk at this conference
\cite{Bijnens:2013zc}. The form of its systematic low-energy expansion is governed
by the symmetry of underlying QCD. This symmetry, called chiral, acts
independently on left and right helicity parts of quark fields. Having $N$
flavours of these quarks the symmetry pattern is $SU(N)_L \times
SU(N)_R$. The property of the vacuum of such a system leads to the spontaneous
symmetry breaking down to the vector subgroup $SU(N)_V$. The consequence of
this symmetry breakdown is the appearance of massless Goldstone boson modes in
spectrum. Due to the small but non-zero mass of these $N$ flavour quarks the
massless Goldstone bosons will be also massive. In practise we can talk about
two possible cases $N=2$ (for $u$ and $d$ quarks) and $N=3$ (for $u$, $d$ and
$s$). At this point everything seems to be prepared for a systematic
construction
of ChPT Lagrangian which would describe the low-energy dynamics of Goldstone
bosons, i.e. three pions for $N=2$ or altogether eight states (pions, kaons and
eta) for $N=3$. However, there is a problem. The number of these particles
would be strictly even in any interaction vertex and it would not be possible to
describe e.g.
$KK\to3\pi$ or $\pi^0\to\gamma\gamma$, i.e. well-established and non-negligible
processes.
The problem lies in the axial current, as this deserves more detailed study due
to the anomaly, in fact two anomalies. One is connected with the so-called
$U(1)_A$ problem and will not be considered here. The other one, chiral anomaly,
connected with electromagnetic interaction is responsible for the existence of
odd-intrinsic parity sector. Its form in ChPT at LO, so-called WZW, can be found
in \cite{chptw} (see also references therein) and for NLO in \cite{chptw2}.

\section{Anomalous processes}

Having formally the Lagrangian at NLO for anomalous processes let us present
here a short overview of its  applicability in the recent ten-year period.
The possible determination of anomalous low energy constants from phenomenology
was established in diploma thesis of O.~Strandberg \cite{Strandberg:2003zf} (see also \cite{Jiang:2010wa}).
One of the most important anomalous process, the decay of pseudoscalar boson
into two photons will be discussed separately in the next subsection. Its
tightly connected process, so-called Dalitz decay, $\pi^0\to e^+e^-\gamma$
was revisited using the Lagrangian of NLO in \cite{Kampf:2005tz}. LECs were set
using the lowest meson dominance model established in \cite{Knecht:2001xc}.
Processes involving kaon, namely radiative kaon decays,
$K_{\ell2\gamma},K_{\ell3\gamma},K_{\ell4}$ and the contributions of the
anomalous sector was considered e.g. in~\cite{kaons}. For an overview not only
on these but all kaon decays see also \cite{Cirigliano:2011ny}.
For the discussion on the relevance of the anomalous sector for radiative pion
decays see \cite{radpide}. The effect of the anomaly in $\pi\gamma\to\pi\pi$
was studied recently \cite{Hoferichter:2012pm} (see also \cite{Bijnens:2012hf}).
The anomaly is important also for processes with $\eta$ \cite{arteta}.
As a further possible option to test anomalous contribution one can use
hadronic tau decays. These were studied e.g. in \cite{hadtaus}. However, the
effect of the odd-intrinsic-parity sector was included only via resonances. The
systematic study of connection between low-energy odd sector and underlying
resonances was subject of \cite{Kampf:2011ty} (see also
\cite{RuizFemenia:2003hm}).

\subsection{$\pi^0\to\gamma\gamma$ and meson decay constant}
This process is described by the WZW Lagrangian and at leading order is thus
fixed
entirely by the anomaly and is parameter free. It is interesting to notice that
there are no chiral logarithms at one loop level \cite{piggnlo}.
This motivated the calculation at next-to-leading order \cite{Kampf:2009tk}.
The result can be further rewritten using the modified counting, which is
\begin{equation}
m_u,\, m_d \sim O(p^2),\qquad m_s \sim O(p)\,.
\end{equation}
It enables to reduce the numbers of odd-intrinsic LECs to two ($C^W_7$ and
$C^W_8$). The amplitude is
\begin{align}
&& T_{(LO+NLO)_+}= {1\over F_\pi}\Bigg\{
{1\over 4\pi^2}
-{64\over3}m_\pi^2C^{Wr}_{7}
+{1\over 16\pi^2} {m_d-m_u\over m_s}
\Big[ 1-{3\over2}{m_\pi^2\over16\pi^2F_\pi^2}L_\pi \Big]
\nonumber\\
&&\phantom{T_{NLO+}}
+32B(m_d-m_u)\Bigg[  {4\over3}C^{Wr}_{7}
+ 4C^{Wr}_{8}\Big(1 -3  {m_\pi^2\over16\pi^2F_\pi^2}L_\pi \Big)
\\
&& \phantom{T_{NLO+}}
-{1\over16\pi^2 F_\pi^2} \Big( 3L_7^r +L_8^r
-{1\over512\pi^2}(L_K+{2\over3}L_\eta) \Big)  \Bigg]
-{1\over24\pi^2} \left( {m_\pi^2 \over16\pi^2 F_\pi^2}L_\pi\right)^2
\Bigg\}\nonumber\ ,
\end{align}
where $L_\pi$, $L_K$ and $L_\eta$ represent chiral logarithms for pion, kaon and
eta respectively, e.g. $L_\pi=\log\frac{m_\pi^2}{\mu^2}$.
Using the existing phenomenological information we come to the following
prediction
\begin{equation}
\Gamma_{\pi^0\to\gamma\gamma} = (8.09 \pm 0.11)\,\text{eV}\,,
\end{equation}
which is in a very good agreement with the existing most precise measurement
by PrimEx collaboration \cite{Larin:2010kq} (see also contribution of Rory
Miskimen at this conference).
Naturally the most important phenomenological input is the pion decay constant
$F_\pi$. Our best estimate led to
\begin{equation}\label{Fpihat}
F_\pi = 92.22 \pm 0.07\;\text{MeV}\,.
\end{equation}
Determination of this value is from the weak charged pion decay, based on the
standard $V-A$ interaction. At this point one can ask how reliable is this
connection. One can assume some variant of SM, e.g. as proposed in
\cite{Bernard:2007cf} by the existence of right-handed current. Thus in
principle we should distinguish $\hat F_\pi$ obtained from
$\pi^+\to\mu^+\nu(\gamma)$ and $F_\pi$ which is the parameter in ChPT.
Schematically
\begin{equation}
F_\pi^2 = \hat F_\pi^2 (1+\epsilon),\qquad \epsilon \sim
\frac{V_R^{ud}}{V_L^{ud}}\,,
\end{equation}
where we have used notation of the above mentioned right-handed currents, but
obviously the meaning can be more general. Now we can turn around the use of the
theoretical $\pi^0\to\gamma\gamma$ decay-width formula. Then $F_\pi$ is an
unknown parameter we want to fix from a $\pi^0$ experiment. Using the most
precise experimental determination for
$\pi^0$ decay width \cite{Larin:2010kq} we can obtain
\begin{equation}
F_\pi = ( 93.85 \pm 1.3\,[\text{exp}] \pm 0.6 \,[\text{theory}] )\;\text{MeV}\,.
\end{equation}
Comparing with the numerical value for $\hat F_\pi$ (\ref{Fpihat}), one can
estimate $\epsilon \approx (3 - 4)\%$, which implies roughly $1\sigma$
significance for the right-handed currents (or any model beyond SM).

\section{Leading logarithms}

In the previous section we have shown the use of ChPT for the real two-loop
calculation and discuss one aspect of the importance of such calculation for
phenomenology. In this section we will focus on another aspect
connected with theoretical ChPT calculations. It is based on
arguments introduced in \cite{LL1} and further developed in \cite{LL2} and
\cite{Bijnens:2012hf}. Short introduction and overview can be also found in
Section 6 in \cite{Bijnens:2013zc} of these proceedings.

The leading logs (LL), i.e. the logarithms with highest power at the given
order (i.e. LL$^1$ at one-loop order, LL$^2$ at two-loop order and so on), are
of special interest. They are parameters-free and can be calculated, in
principle, to all orders from one-loop diagrams only. Their eventual
resummation then leads to the same effect as is the concept of running coupling
constant for the renormalizable theories. In the above mentioned articles the
procedure of the one-loop calculations was automatized and high orders of LL
were presented. This was done for a general $O(N+1)/O(N)$ model which coincides
with $SU(2)$ for $N=3$. Another possible group structure $SU(N)\times
SU(N)/SU(N)$ with the direct correspondence to ChPT for both $N=2$ and $N=3$
(though only for the equal-mass case) is under development \cite{SUNLL}.

\begin{figure}
\begin{center}
\begin{picture}(0,0)
\put(230,-10){LL$^i$}
\put(-5,135){\%}
\end{picture}
\includegraphics[width=0.6\textwidth]{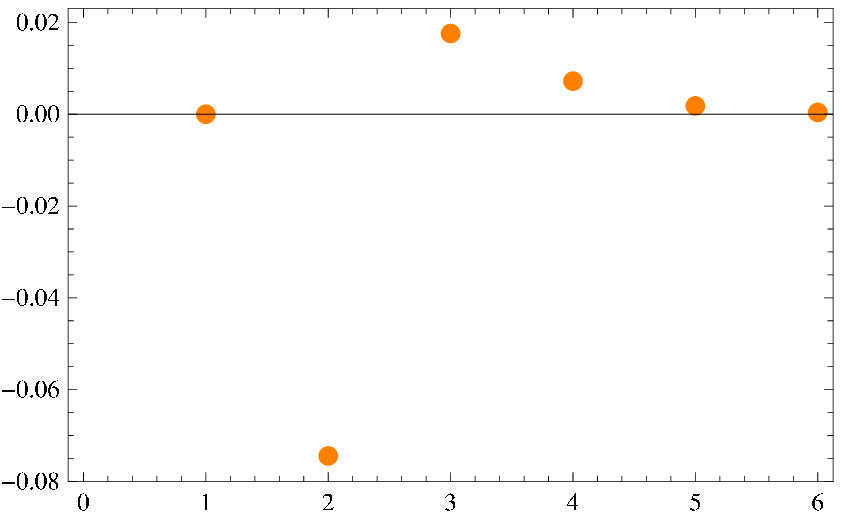}
\end{center}
\caption{The leading logarithm contribution at individual orders in percent of
the leading order for $\Gamma(\pi^0\to \gamma\gamma)$.}
\label{figpigg}
\end{figure}

It is clear that main motivation for LLs is theoretical as it can hint to some
deeper understanding of calculation within effective field theories.
However, their application can be also phenomenological. As already stated,
they are not proportional to some unknown parameters. We can define them using
the lowest-order parameters of the Lagrangian as
\begin{equation}
L = \frac{M^2}{16\pi^2 F^2} \log \frac{\mu^2}{M^2}\,,
\end{equation}
or using the physical mass and decay constant
\begin{equation}
L_\pi = \frac{M_\pi^2}{16\pi^2 F_\pi^2} \log \frac{\mu^2}{M_\pi^2}\,.
\end{equation}
The high orders of LL of several quantities relevant for ChPT were already
calculated. General notation of the expansion is as
follows
\begin{equation}
 O_\text{phys} = O_0 (1 + \sum_i a_i L^i),\qquad O_\text{phys} = O_0 (1 + \sum_i
c_i
L_\pi^i)\,.
\end{equation}
More precisely, it was studied: expansion of mass and decay constant,
vacuum expectation
value, vector and scalar formfactor, $\pi\pi$ scattering and in the forthcoming
article \cite{SUNLL} also $\gamma\gamma\to\pi\pi$ process. Concerning the
anomalous processes, LL for the quantities connected with $\pi\gamma\to\pi\pi$
and $\pi^0\to\gamma\gamma$ were calculated up to the 6th and 7th loop
respectively. The calculation of the latter can be used to demonstrate the
convergence of the perturbative calculation of $\pi^0$ decay width. As stated in
the previous section already the leading order is in excellent agreement with
experimental measurements. This is a signal of a good convergence of the
perturbative series and can be graphically demonstrated using the LL
calculation (Fig.\ref{figpigg}). 

Let us conclude this section with a remark on automatized calculation of LL and its development in time.
Following the articles in \cite{LL2} one can notice that first results for
expansion for physical mass and physical decay constant were presented up to 5th loop order (in $O(N+1)/O(N)$ model).
Approximately one or two years later it was
possible to add one order more \cite{Bijnens:2012hf}. Now we are again after
similar time interval and we can present the highest order obtained for these
quantities, the 7th loop order. Corresponding tables (Tables 1--4 in
\cite{Bijnens:2012hf}) can be thus extended by the following formulae, first for
the \underline{physical meson mass}:
\begin{align}
 a_7 &=  1098817478897/8573040000 - 286907006651/1428840000\,N \notag\\
 & + 4533157401977/11430720000\,N^2 - 1986536871797/3429216000\,N^3\\
 & + 436238667943/762048000\,N^4 - 7266210703/21168000\,N^5 + 
 99977/896\,N^6 - 15 N^7\notag
\end{align}
and
\begin{align}
 c_7 &= 1516884225443/34292160000 - 315684201397/2857680000\,N \notag\\ 
   & +1125614672041/15240960000\,N^2 + 174975088027/3429216000\,N^3 \\
   & - 38300257501/609638400\,N^4 + 1140619717/56448000\,N^5
      -355/21504\, N^6-  33/2048\,N^7\,.\notag
\end{align}
Similarly for the \underline{physical decay constant} we have obtained
\begin{align}
a_7 &=  - 1560715223869/182891520000 + 15484111239353/548674560000\,N \notag\\
 &- 447910156369/7315660800\,N^2 + 63001235724859/548674560000\,N^3  \\
  &- 178022410793/1219276800\,N^4 + 9048751829/84672000\,N^5 
- 8993/224\,N^6 + 6 N^7 \notag
\end{align}
and
\begin{align}
&c_7 = -5072297701349/182891520000 + 29611209759293/548674560000\,N \notag \\
 &+750961017899/36578304000\,N^2- 62450118717821/548674560000\, N^3\\
&+560319794369/6096384000\,N^4 - 2970744467/112896000\,N^5 
+44777/35840\,N^6 + 429/2048\,N^7\notag
\end{align}
We have checked the correct large $N$ behaviour as calculated in \cite{LL2}.
Note that setting $N=3$ one can obtain the results relevant for the two-flavour
ChPT.


\end{document}